\documentclass[sigplan,nonacm,10pt]{acmart}

\usepackage{listings}
\usepackage{enumitem}
\usepackage{algpseudocode}
\graphicspath{{figs/}{./}}
\usepackage{microtype}

\newcommand{\code}[1]{\texttt{#1}}
\DeclareRobustCommand{\_}{\textunderscore\allowbreak}

\hypersetup{
  colorlinks=true,
  linkcolor=blue,
  citecolor=blue,
  urlcolor=blue
}

\lstdefinelanguage{JSON}{
  basicstyle=\ttfamily\footnotesize,
  morestring=[b]",
  morecomment=[l]{//},
  stringstyle=\color{red!60!black},
  commentstyle=\color{gray!80},
  showstringspaces=false
}
\begin{document}

\title{LearnedCache: eBPF-Integrated Perceptron-Based Eviction Policies for the Linux Page Cache}

\author{Zejia Qi}
\affiliation{%
  \institution{}
  \city{}
  \country{}
}

\renewcommand{\shortauthors}{Qi}

\begin{abstract}
Any device that runs Linux uses the Linux page cache, a central pillar in OS and application performance, serving to
reduce extraneous disk access. Many page cache eviction policies have been developed but remain bound by the
rigidity of heuristics. Promising research
has been done on neural cache eviction policies, but only in the field of user-space applications such as CDNs. We present LearnedCache, a set of
machine-learning-based page cache eviction policies that run live inside the
Linux kernel through the \code{cache\_ext} eBPF framework. LearnedCache trains a
lightweight perceptron per workload on kernel trace data collected using eBPF to
predict whether eviction candidates will be reused within a bounded horizon and
deploys it through two eBPF policies. The resulting policies
consistently compete with and outperform both external heuristic policies and
the kernel's own, improving application throughput by up to 44\% versus standard kernel policies.

\end{abstract}


\maketitle

\section{Introduction}
Linux serves as a bedrock for modern digital infrastructure, running across an
immense plethora of devices. Yet, with this dominance, the adaptation of custom
page cache algorithms beyond those shipped in the kernel itself has been slow to
follow, creating a longstanding bottleneck that, if minimized, would potentially
offer large performance gains. Dating back to 1981, Michael Stonebraker
underscored the need for eviction policies beyond one-size-fits-all
LRU/MRU-derived algorithms, arguing that these could not accommodate the growing
variety of database workloads at the time~\cite{stonebraker1981}. In the modern
day, this fact still holds---the Linux kernel still uses an LRU derivation,
namely Multi-Generation LRU, across the even wider range of workloads it
handles~\cite{corbet2021mglru}. One can attribute this lack of adaptation to
accessibility---up until recently, there existed no high-performance solution to
insert a custom eviction policy into the page cache without requiring kernel
patches, holding a high skill barrier to entry.

eBPF introduces a way for sandboxed user-written programs to be algorithmically
verified for unsafe behavior before they are run in kernel space, opening new
venues for custom page cache eviction
policies~\cite{ebpf_io, gbadamosi2024ebpfruntimelinuxkernel}. \code{cache\_ext}, created
near the end of 2025, introduces a framework that allows user-developed eBPF
programs to interface with the provided custom kernel to create easily swappable
policies with near-kernel-level performance~\cite{cache_ext}.
However, the policies offered by \code{cache\_ext} and those that remain the de facto
standard remain mostly limited to heuristics---still being variations of LRU,
MRU, and FIFO, albeit with newer variations such as the probability-based
LHD coming into play~\cite{beckmann2018lhd}. It is also shown in
\code{cache\_ext} that no single policy performs well across all
workloads~\cite{cache_ext}.

\begin{sloppypar}
A lightweight machine-learning-based eviction policy would theoretically be able
to adapt to any workload and outperform traditional policies, creating a highly
specific policy for any workload given trace data to train on. An ML-based
eviction policy would also save time and effort in creating tailored algorithms
for the wide variety of workloads present in the cloud and other infrastructure.
Works have attempted to create machine learning solutions for cache eviction:
ML-CLOCK outlines a possible perceptron-based eviction policy for page caches in
a ``trace-driven simulation'' but does not publish the specific details or the
simulation used~\cite{ml_clock}. Yang et al.\ offer performant ML-based eviction
policies by adding an ML model on the tail end of a heuristic
algorithm~\cite{yang2023_learned_cache}, and Song et al.\ in HALP introduce a
trained eviction policy for the YouTube CDN, achieving a 9.1\% reduction in byte
miss ratio~\cite{song2023_halp}.
\end{sloppypar}

\begin{figure}[t]
\centering
\includegraphics[width=0.95\columnwidth]{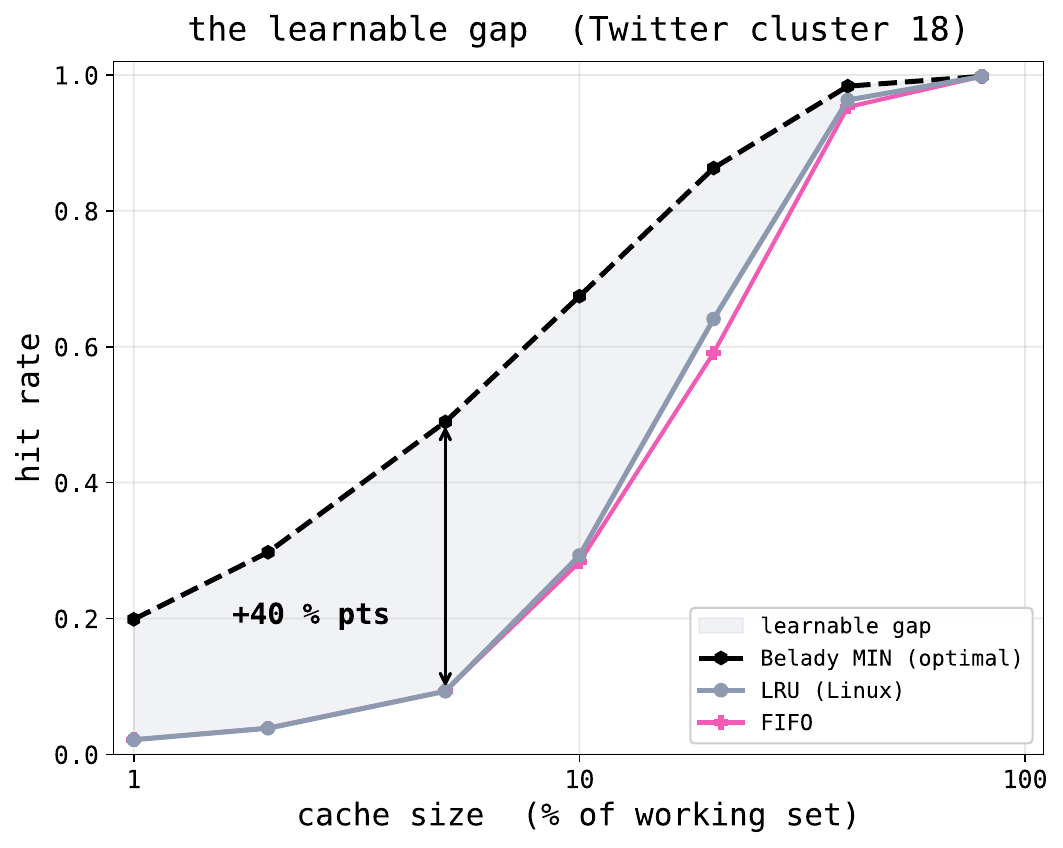}
\caption{The learnable gap: across cache sizes on a representative Twitter trace
(cluster~18), the kernel's LRU and FIFO trail the Belady-optimal hit rate by up
to ${\sim}40$ percentage points (shaded).}
\Description{A hit-rate versus cache-size plot in which LRU and FIFO fall well
below the Belady-optimal curve; the shaded region between them is labelled the
learnable gap.}
\label{fig:opportunity}
\end{figure}

Although ML-driven cache eviction policies show potential, no research thus far has trained one and then
placed it into the Linux page cache, most likely due to the tooling constraints.
What has been presented is that lightweight models that augment the
decisions of heuristics are proven to be effective in other application contexts,
and LearnedCache extends upon this with the belief that the same can be done
in the kernel itself, and that workloads inherently hold reuse signal in simple features---for example, few workloads 
access pages completely randomly, instead having underlying algorithms
or distributions that can be exploited for gain. Figure~\ref{fig:opportunity}
quantifies this opportunity: on data collected from rerunning Twitter cache traces~\cite{yang2020twitter}, the kernel's
LRU and FIFO trail the Belady-optimal hit rate by as much as 40 percentage
points.

LearnedCache embeds single perceptrons
over simple derived features such as access deltas and explores two different policies
using these models to aid with heuristic eviction decisions. These policies
run live in the Linux kernel via the
\code{cache\_ext} eBPF harness and consistently compete with and outperform both
\code{cache\_ext} and kernel heuristic policies, improving application
throughput by up to 44\% versus kernel LRU and MGLRU.

The goal of our work, thus, is to demonstrate the following through LearnedCache:
\begin{itemize}[leftmargin=*]
\item Accurate and performant neural models can be trained using kernel trace
data of varying workloads.
\item eBPF-driven implementation of policies utilizing such models is possible
with the help of \code{cache\_ext}.
\item These models and policies can still show significant improvement over
heuristic policies in a kernel context.
\end{itemize}

\section{Background}
A cache is a high-speed, temporary transport-layer storage that stores copies of
frequently accessed data to improve access times. Operating systems use a page
cache to store fixed-size chunks (usually of 4\,KiB or 4096 bytes), known as
pages, of files on disk into much faster random access memory (RAM) for faster
repeated reads and writes~\cite{silberschatz2018operating}. Because the amount of
free RAM a computer has is (in the majority of instances) far less than the
amount of disk space, reads of multiple files from disk will fill the page cache
up quickly, requiring eviction, or removal, of pages from the cache. The
algorithm that decides which pages to evict from the cache is called an eviction
policy.

\subsection{Folios}
Linux has recently begun transitioning to folios in place of
pages~\cite{corbet2021folios}. A folio is either a zero-order page, or a
group of pages contiguous in memory that can be treated like a singular larger
page. The terms folio and page will be used interchangeably, and all folios
mentioned are equivalent to a single page.

\subsection{\code{cache\_ext}}
\code{cache\_ext} is a recent eBPF-based framework designed to enable developers
to implement page cache eviction policies in the kernel with minimal
overhead~\cite{cache_ext}. User-created eBPF programs mount onto a series of hook
points provided by the \code{cache\_ext} kernel and manage pages through their
own algorithms and lists. \code{cache\_ext} provides a built-in trusted data
structure for storing lists of pages for convenience, along with helper
functions to traverse and modify them. All \code{cache\_ext} programs can be
bounded to listen only to a single file directory and to programs within a
cgroup, which is utilized by LearnedCache for reproducibility~\cite{cgroupv2}.

\section{Design and Implementation}
In this section, we present the design used for LearnedCache. The project can be
split into three main pipeline components:
\begin{itemize}[leftmargin=*]
\item The eBPF tracer, which collects data from workload runs to feed
the model.
\item The model, which is trained to predict whether eviction
candidates will be reused within a certain timeframe.
\item The policies, which run the model using eBPF to produce eviction decisions
in real time during workload execution.
\end{itemize}

\subsection{Tracer Design}
We will now describe the design of the tracer. \code{cache\_ext} programs are
exposed to a \code{struct\_ops} interface, creating functions that run on cache
eviction requests, eviction itself, insertion, and access~\cite{cache_ext}.
Notably, access hooks into the kernel function \code{folio\_mark\_accessed},
which is called by the kernel to mark a folio as recently
accessed~\cite{linux_folio_mark_accessed}. Eviction and insertion hook onto
\code{\_\_filemap\_remove\_folio} and \code{\_\_filemap\_add\_folio},
respectively~\cite{linux_filemap}. Because \code{folio\_mark\_accessed} is inherently used as a signal
that a folio is active rather than a function that is called on each access, it
is only possible to treat it as a roughly representative subsample of all
accesses, which works fine for modeling.

Given this accessibility, tracing for data collection is done using a custom
policy that, in addition to acting as a simple FIFO policy with LFU
oversampling, logs eviction, access, and insertion events back to userspace via a
BPF ring buffer. The userspace loader then streams these logs into compact binary
logfiles for later reading during model training.

Initial feature sets that used absolute access counts were discarded because they
contained little meaningful information and were temporally dependent. Features
such as the absolute number of accesses to a page could not be used because they 
were dependent on tracer initialization and therefore could not be generalized to other
runs at arbitrary times. Additionally, all features must be computable during eviction to
meaningfully compare model predictions. After iteration and testing, the
following final feature set was settled upon, constructed for each page from
its own access and eviction history.
\begin{itemize}[leftmargin=*]
\item Delta time between last accesses of page.
\item Delta time between last 2 accesses of page (i.e.\ second and third to last).
\item Delta time between last access of inode.
\item Delta time between last 2 accesses of inode (i.e.\ second and third to last).
\item Distance from last access within file.
\item File size in pages.
\item Relative exponential moving average for page accesses.
\item Relative exponential moving average for inode accesses.
\item Time from last access to eviction.
\end{itemize}

The two relative exponential moving averages track access frequency at the page
and inode granularity. Each maintains a score $S$ that, on every access, decays
with a half-life $T = 1\text{\,s}$ before the current access is counted:
\begin{equation}
S \leftarrow S \cdot 2^{-\Delta t / T} + 1,
\label{eq:ema}
\end{equation}
where $\Delta t$ is the time since the previous access. The score is held as a
fixed-point integer (scaled by $1000$) so it can be maintained without
floating-point operations in eBPF: for $\Delta t \ge T$ the exponent is floored
to an integer number of half-lives ($2^{-\lfloor \Delta t/T\rfloor}$, a
bit-shift), for $\Delta t < T$ the decay factor is
approximated linearly as $1 - \tfrac{1}{2}(\Delta t / T)$, and decays beyond ten
half-lives are clamped to zero.

These states can be represented as a function solely of the timeline of page
accesses and evictions, decoupling them from insertion events. Concretely, this
state is maintained in two BPF hash maps that are updated at every access event:
one keyed by the (device, inode, offset) tuple identifying an individual page,
which records per-page timing (first, previous, and last access times and their
deltas), the last file offset, file size, and an access-frequency counter; and
one keyed by (device, inode) identifying a file, which records per-inode timing
deltas, a hotness EMA, and the last accessed offset. At every access this data
is transmitted to the BPF userspace loader via ring buffers.

\subsection{Model Design}
We will now outline the model design, which is trained offline per workload using
logs collected by the tracer. The following architecture was chosen for its low
computational cost, which is needed to run over multiple pages in the cache upon
every kernel eviction.

The model is implemented as a single perceptron with a bias term on the
previously described feature set. For a feature vector $\mathbf{x}$, learned
weight vector $\mathbf{w}$, and bias $b$, it predicts
\begin{equation}
p = \sigma(\mathbf{x} \cdot \mathbf{w} + b),
\label{eq:perceptron}
\end{equation}
where $p$ is the folio's predicted reuse probability and $\sigma$ is the logistic
sigmoid. Training labels are constructed for each access
and its corresponding next eviction event, answering the following question:
``Will this page get reused within a set horizon of the next eviction?'' This
``reuse horizon'' is calculated as the average time a page stays in the cache, or
the average time it would take for the cache to fully turn over under a FIFO
policy. During training, the model is given a sigmoid activation function; during
evaluation time, raw logits are used.

Before feeding into the perceptron, each training row is discretized using
quantile-based binning (with a maximum of 10 bins) and then one-hot encoded. It
should be noted that, at times, the distribution of features does not allow for
discretization to the maximum number of bins. Although this increases decision
overhead, it must be done to induce nonlinearity in the model, giving
it the power to generalize to more relationships between features and reuse
labels.

After a model is trained on the access and eviction logs from a workload run, the
discretizer's interior edges and the model weights, quantized to integers, are
exported as a JSON file. This is eventually parsed by the
model policy loader and saved to eBPF maps; the process of model inference
is described in greater detail in
Section~\ref{subsec:policy}.

\subsection{Policy Design}
\label{subsec:policy}
\begin{figure}[t!]
\centering
\includegraphics[width=0.95\columnwidth]{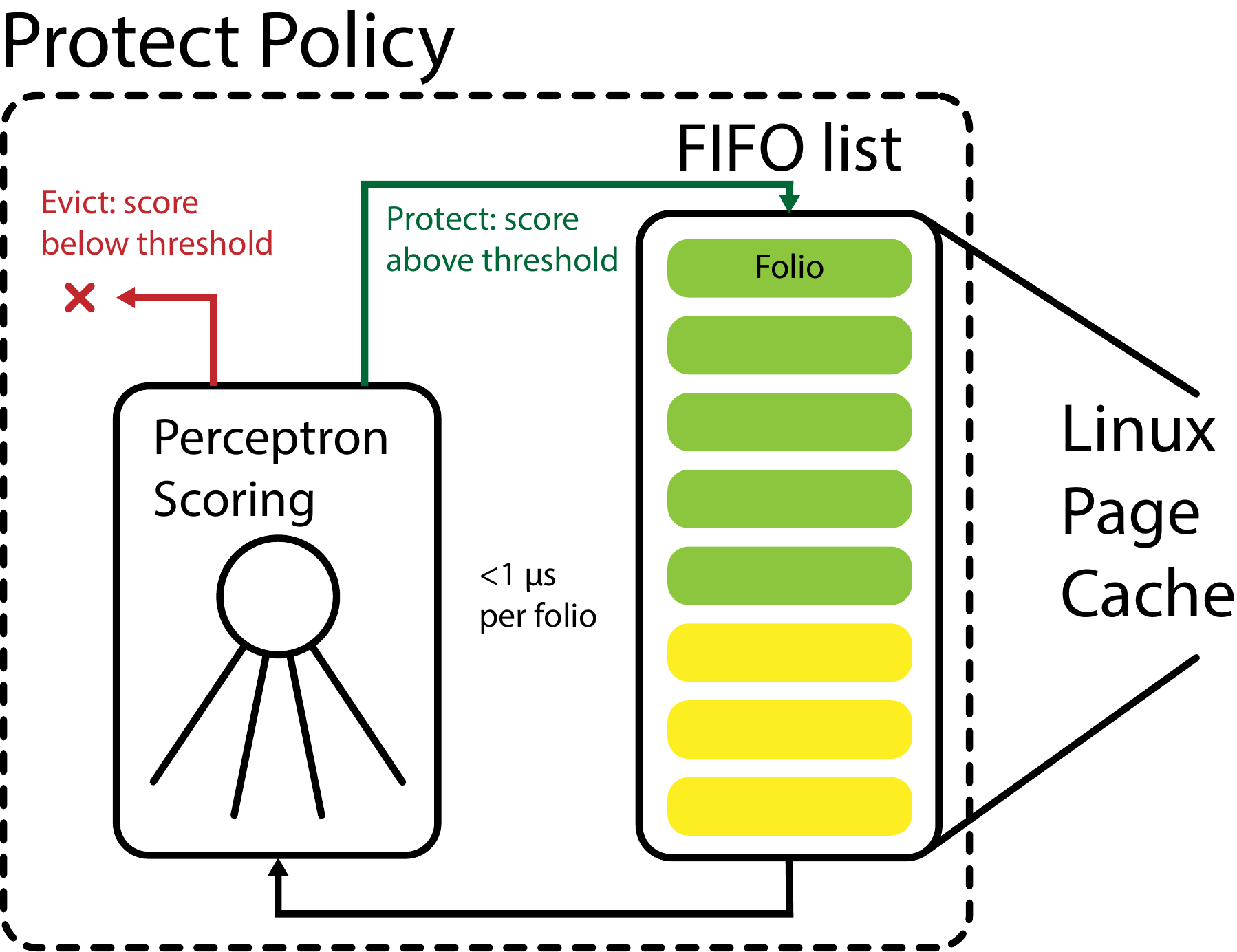}
\caption{The \code{ml\_protect} policy.}
\Description{Schematic of the protect policy scoring the tail folio of a FIFO
list with the perceptron and either evicting it or rotating it to the front.}
\label{fig:protect}
\end{figure}

\begin{figure}[t!]
\centering
\includegraphics[width=0.95\columnwidth]{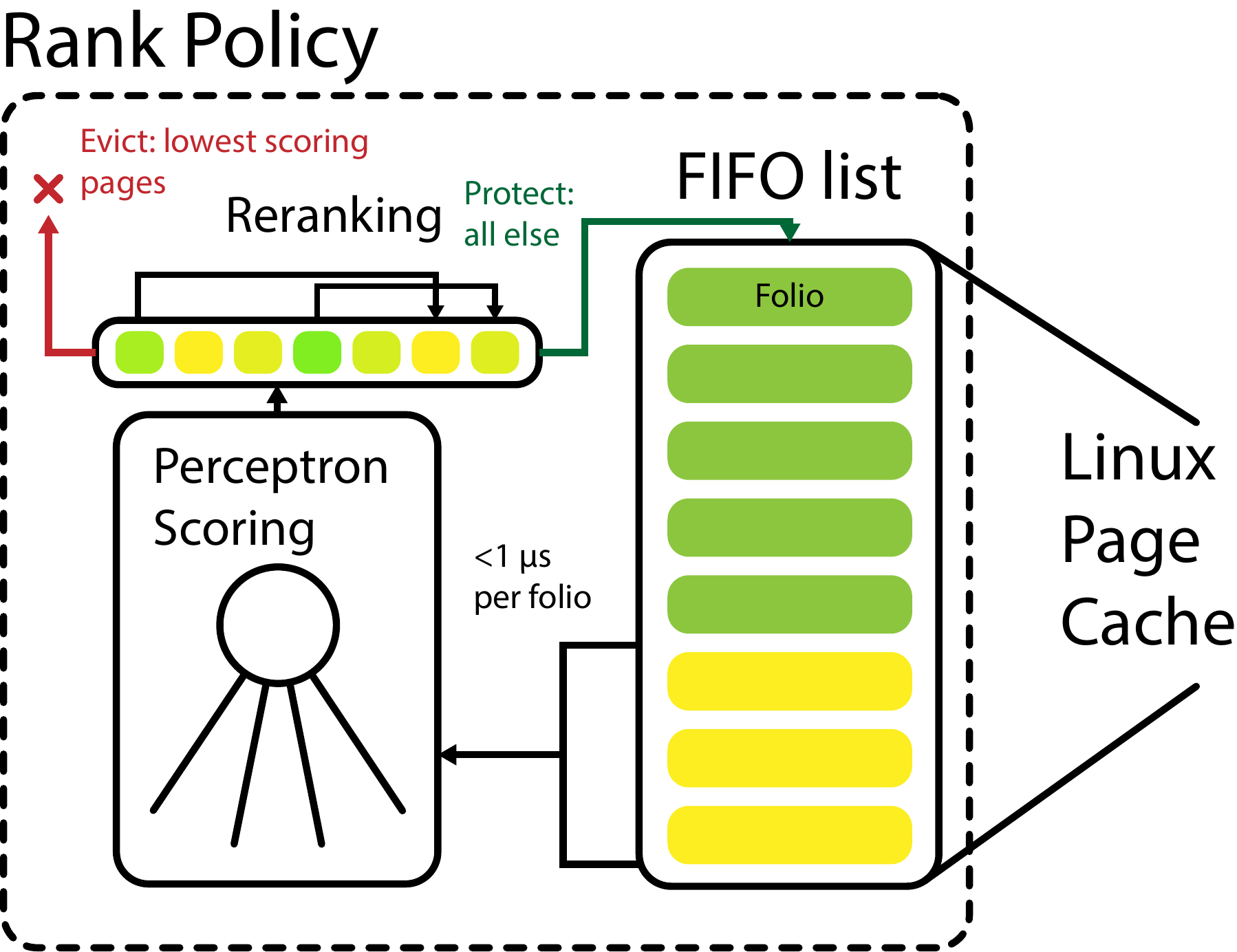}
\caption{The \code{ml\_rank} policy.}
\Description{Schematic of the rank policy reranking an oversampled batch of
folios from a FIFO list with the perceptron and evicting the lowest-scoring.}
\label{fig:rank}
\end{figure}

We will now explore the design of two policies that utilize the model in
differing ways, shown in Figures~\ref{fig:protect} and~\ref{fig:rank}.

\begin{itemize}[leftmargin=*]
\item The first is a relatively low-overhead policy, referred to as
\code{ml\_protect} (Figure~\ref{fig:protect}). It is implemented as a
classical FIFO policy, except each folio at the tail is evaluated by the model
and moved back to the front if found to have a high likelihood of reuse (logit above a set threshold, usually 0);
otherwise it is evicted.
\item The second is a policy with a higher cost, referred to as \code{ml\_rank}
(or \code{ml\_ranking}; Figure~\ref{fig:rank}). It is also implemented
as FIFO, but upon an eviction request it \emph{oversamples}: it reads $n$ times
more pages off the tail of the list than it needs to evict and considers them as
a group rather than one at a time, where $n$ is an adjustable oversampling
factor. It then ranks each page by its probability of reuse, running model
inference on each. The
lowest-scoring pages are then evicted. Because of this oversampling, this policy
generally makes better eviction decisions than \code{ml\_protect}.
\end{itemize}

While the ways the model is invoked differ between these, the core evaluation
logic is the same. Model weights and discretizer edges are stored in eBPF array
maps. The feature set is fixed at nine entries and bins are capped at ten:

\begin{lstlisting}[language=C]
#define NUM_MODEL_FEATURES 9
#define MAX_BINS 10
enum model_features {
  PD  = 0, // page delta t
  SZ  = 1, // size
  // ... other features
  TSA = 8, // time since last access at eviction
};
\end{lstlisting}

Three eBPF array maps, each with \code{NUM\_MODEL\_FEATURES} entries, hold the model:
\code{n\_bins\_map} (value \code{\_\_u8}) stores the number of bins---and
therefore weights---for each feature; \code{bin\_edges\_map} (value
\code{\_\_u64[MAX\_BINS]}) stores bin edges for each feature; and
\code{nn\_weights\_map} (value \code{s64[MAX\_BINS]}) stores per-bin integer
weights. A representative definition is shown below; the other two maps differ
only in their value type.

\begin{lstlisting}[language=C]
struct {
  __uint(type, BPF_MAP_TYPE_ARRAY);
  __uint(max_entries, NUM_MODEL_FEATURES);
  __type(key, __u32);
  __type(value, s64[MAX_BINS]);
} nn_weights_map SEC(".maps");
\end{lstlisting}

One extra single-entry map, \code{model\_meta\_map}, is used by
\code{ml\_protect} to store the (quantized) bias and the protection logit
threshold:

\begin{lstlisting}[language=C]
struct model_meta {
  s64 bias;       // quantized by weight_scale
  s64 threshold;
};
\end{lstlisting}

Below is the pseudocode for model evaluation of a single feature, assuming the
array \code{raw\_features} holds the calculated non-discretized feature values and
\code{score} accumulates the folio's running score.

\begin{algorithmic}[1]
\Procedure{ProcessFeature}{f, raw\_features}\Comment{accumulate into score}
  \State n $\gets$ n\_bins[f] \Comment{number of bins for feature f}
  \If{1 $\le$ n $\le$ MAX\_BINS}
    \State b $\gets$ \Call{Discretize}{raw\_features[f], bin\_edges[f], n}
    \State score $\gets$ score $+$ weights[f][b] \Comment{add the active bin's weight}
  \EndIf
\EndProcedure
\end{algorithmic}

In practice, \code{ProcessFeature} is realized as a C macro,
\code{PROCESS\_FEATURE}, invoked once per feature so that the pass over all nine
features is manually unrolled rather than written as a loop. Each array access
becomes a \code{cache\_ext} array-map lookup guarded by an explicit null
check, and the bin index is clamped before use. The \code{Discretize} step is itself a helper function
that maps a raw value to its bin index by an unrolled comparison against the
feature's bin edges. These steps were found to be the
simplest way to pass program verification without gambling on whether Clang
optimizes away the bounds that the verifier relies on.
The core logic of the model boils down to two steps:
\begin{itemize}[leftmargin=*]
\item Finding which encoding a given raw feature value falls under, then finding
the weight of that specific bin.
\item Summation of each weight found through this process.
\end{itemize}

Because one-hot encoding produces inputs that are either 0 or 1, we are able to
skip explicitly calculating the score for the inactive bins for a feature, and
simply add the weight of the active bin to the final score.

\section{Policy Evaluation and Analysis}
In this section, we present the evaluation methodology and analyze the
performance of the LearnedCache policies. We provide evidence that simple
perceptron-based models can run in eBPF without debilitating overhead, and
demonstrate that ML-based page cache eviction policies are feasible for
application workloads.

\subsection{YCSB}
We first train and evaluate \code{ml\_protect} and \code{ml\_rank} on the suite
of YCSB workloads (excluding the uniform read/write workload), which use a
Zipfian access distribution. A separate model is trained for each workload. These are run on the widely used LevelDB
key--value store~\cite{leveldb}, mirroring the original evaluation done on
\code{cache\_ext}~\cite{cache_ext}. Alongside the two model policies, both kernel
heuristics (Linux LRU and MGLRU) and all \code{cache\_ext} heuristics (FIFO, MRU,
LHD~\cite{beckmann2018lhd}, LFU, and S3-FIFO~\cite{yang2023s3fifo}) are run for
comparison. A reproducible 100\,GiB database is used, with a working cgroup of
10\,GiB. LevelDB is built with the \code{cache\_ext}-induced modification of
using \code{pread} instead of \code{mmap} to improve performance. For each YCSB
workload, a run using the tracer policy was first performed; the model was then
trained offline; and finally, a run with \code{ml\_protect} and \code{ml\_rank}
at four different oversampling factors ($10\times$, $20\times$, $30\times$, and
$40\times$) was performed. The evaluated workloads and their request
compositions are summarized in Table~\ref{tab:ycsb}.

\begin{table}[t]
\centering
\caption{YCSB core workloads evaluated (the uniform read/write workload is
excluded).}
\label{tab:ycsb}
\footnotesize
\begin{tabular}{@{}cll@{}}
\toprule
\textbf{Workload} & \textbf{Type} & \textbf{Request composition} \\
\midrule
A & Update-heavy      & 50\% read, 50\% update \\
B & Read-mostly       & 95\% read, 5\% update \\
C & Read-only         & 100\% read \\
D & Read-latest       & 95\% read, 5\% insert \\
E & Short ranges      & 95\% scan, 5\% insert \\
F & Read-modify-write & 50\% read, 50\% read-modify-write \\
\bottomrule
\end{tabular}
\end{table}

LevelDB is a log-structured merge-tree store~\cite{oneil1996lsm}: writes are first
buffered in an in-memory memtable and then flushed to immutable on-disk sorted
string tables (SSTables) organized into levels. To bound read amplification and
reclaim space from overwritten or deleted keys, a background \emph{compaction}
process continually merges overlapping SSTables from one level into the next,
rewriting their contents~\cite{wisckey}. This compaction, driven by updates, deletions, and writes,
generates I/O that is largely independent of the application's own access
pattern, weakening access frequency as a predictor
of future reuse---an effect that becomes important on the write-heavy workloads
discussed below.

\begin{figure}[t!]
\centering
\includegraphics[width=0.95\columnwidth]{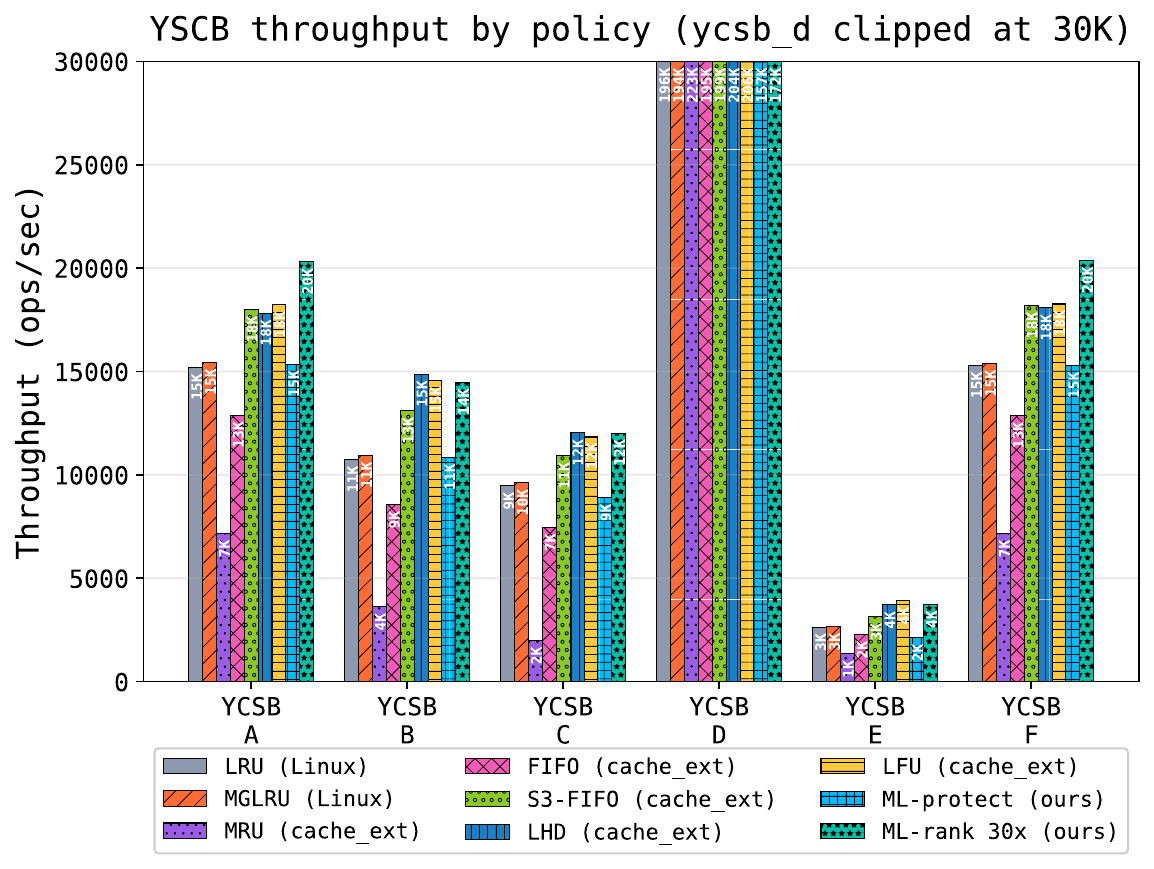}
\Description{Bar chart of YCSB throughput per workload for each policy.}
\caption{YCSB throughput (ops/sec) by policy across workloads.}
\label{fig:ycsb-tput}
\end{figure}

As shown in Figure~\ref{fig:ycsb-tput}, \code{ml\_rank} with an oversampling
factor of 30 consistently exceeds both kernel heuristics in throughput by
significant margins, up to a 40\% increase on YCSB~E over MGLRU. It is only on
YCSB~D that the learned policies struggle, owing to their increased overhead in
such a simple, read-heavy workload. \code{ml\_protect}, on the other hand,
consistently performs slightly worse than the kernel heuristics: it can only
enhance FIFO's decisions, which are poorly suited to YCSB.

\begin{figure}[t!]
\centering
\includegraphics[width=0.95\columnwidth]{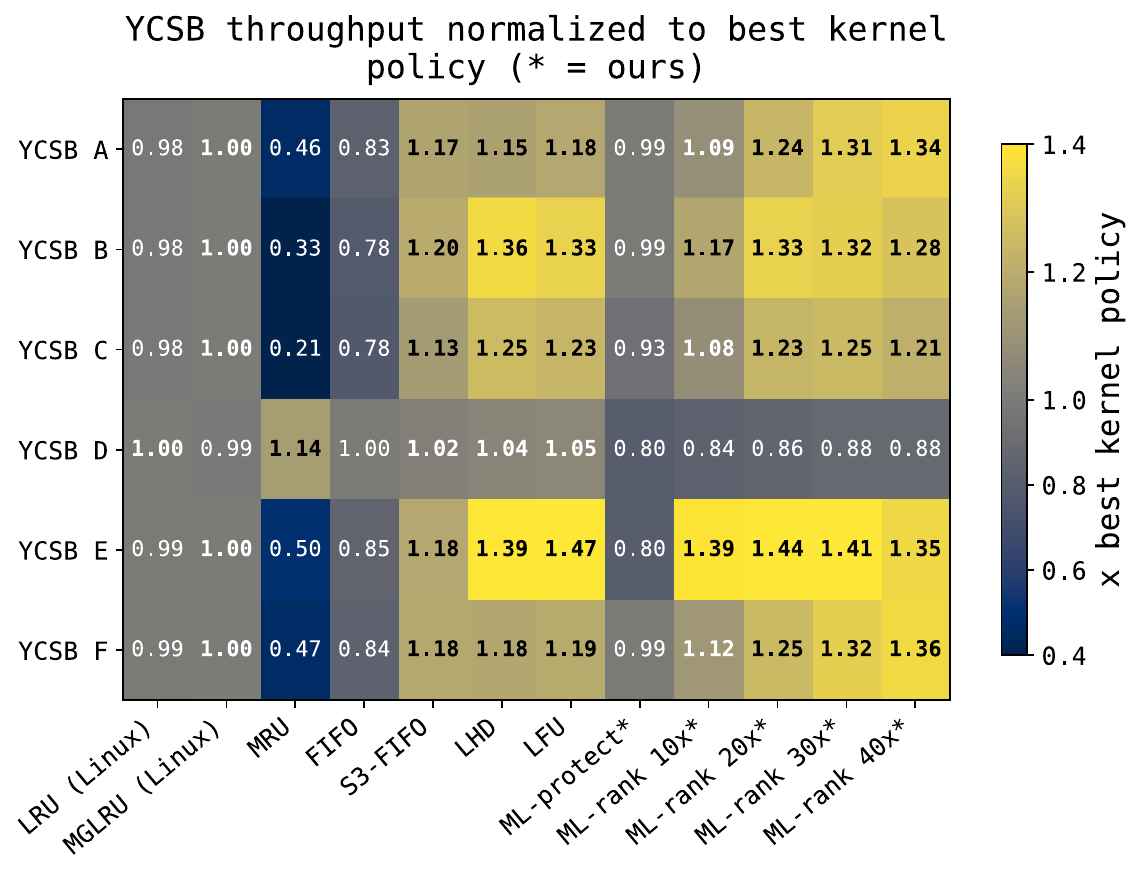}
\Description{Heatmap of YCSB throughput per policy normalized to the best kernel policy.}
\caption{YCSB throughput by policy, normalized to the best kernel policy per
workload (the stronger of Linux LRU and MGLRU; $\times$\,best kernel;
${}^{*}=$\,ours).}
\label{fig:ycsb-heatmap}
\end{figure}

Figure~\ref{fig:ycsb-heatmap} normalizes these throughputs to the best kernel
policy (the stronger of Linux LRU and MGLRU) per workload. Compared to the
heuristics from \code{cache\_ext}, however, the performance gain is more modest. Because LFU and LHD are very good
approximations of the YCSB
Zipfian access distribution, they consistently perform as the top heuristic
policies, tying with or edging out \code{ml\_rank} on YCSB~B, C, and E, while
\code{ml\_rank} beats the top heuristic by a margin of about 13\% on YCSB~A and
F. This can be explained by the fact that A and F are much more write-heavy than
the other workloads, decoupling access frequency as an indicator of page reuse
through LevelDB compaction during writes. Whereas the model generalizes to
approximate an LHD-like policy with added overhead on the other YCSB workloads,
on A and F it is able to distinguish page reuse from extra signals in its feature
set that carry more information about reuse.

\begin{figure}[t!]
\centering
\includegraphics[width=0.95\columnwidth]{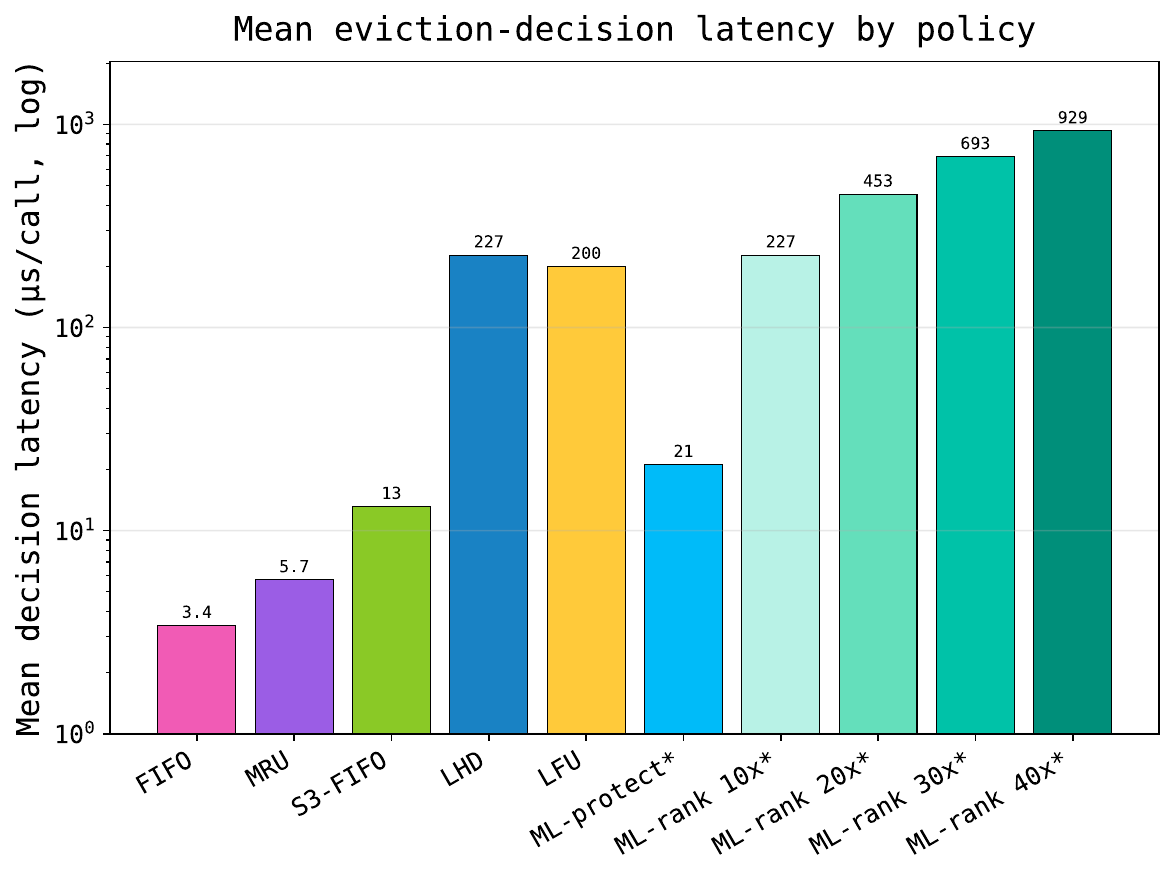}
\Description{Bar chart of mean eviction-decision latency per policy and oversampling factor.}
\caption{Mean eviction-decision latency. Larger oversampling factors incur
substantially greater decision overhead; \code{ml\_protect} is shown on YCSB~A.}
\label{fig:latency}
\end{figure}

The per-factor \code{ml\_rank} columns of Figure~\ref{fig:ycsb-heatmap} bear this
out across oversampling factors: models find more signal in YCSB~A and F, with
the gain in decision quality from discernment between pages dominating the
increased overhead at higher sampling factors, whereas for the other workloads
the opposite happens. From the mean eviction-decision latency
graph (Figure~\ref{fig:latency})---in which \code{ml\_protect} was run on YCSB~A
because of its variable decision latency---it is evident that oversampling by
larger amounts results in much greater decision overhead, substantiating the
previous points.

\subsection{Twitter Traces}
The authors of \code{cache\_ext} find that, in the presence of real-world
workloads, no one heuristic policy dominates all others~\cite{cache_ext}. This is due
to the inherent unpredictability of real workloads: there is no specified
distribution for accesses to follow, nor are there any hardcoded rules for access
patterns. We train and evaluate the LearnedCache policies on production traces
from Twitter's cache workloads~\cite{yang2020twitter}, using the same cgroup
memory allocation as
\code{cache\_ext}: 10\% of the database size. A separate model is trained for each trace workload. Because this induces extreme cache
pressure, \code{ml\_rank} is wholly infeasible due to its substantial overhead,
and only \code{ml\_protect} is evaluated.

\begin{figure}[t!]
\centering
\includegraphics[width=0.95\columnwidth]{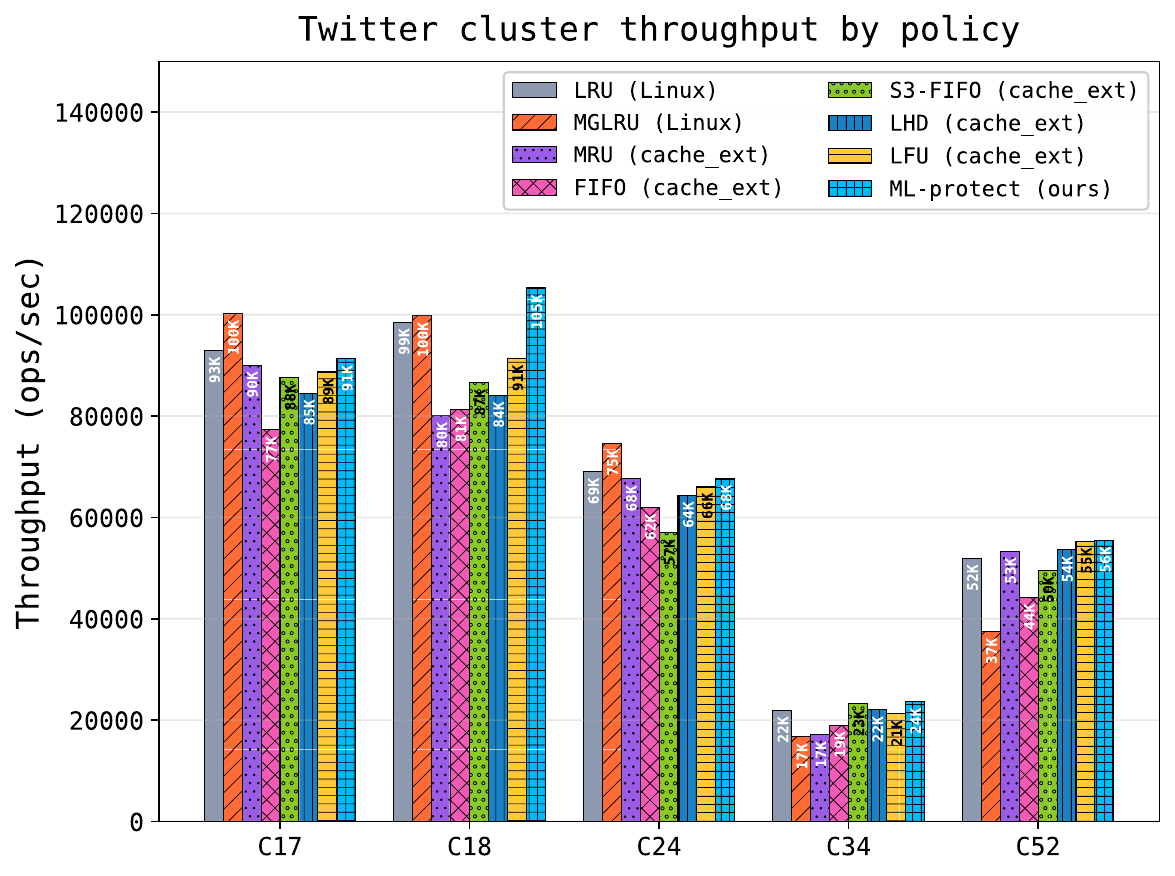}
\Description{Bar chart of Twitter cluster throughput per policy.}
\caption{Twitter cluster throughput (ops/sec) by policy.}
\label{fig:twitter-tput}
\end{figure}

\begin{figure}[t!]
\centering
\includegraphics[width=0.95\columnwidth]{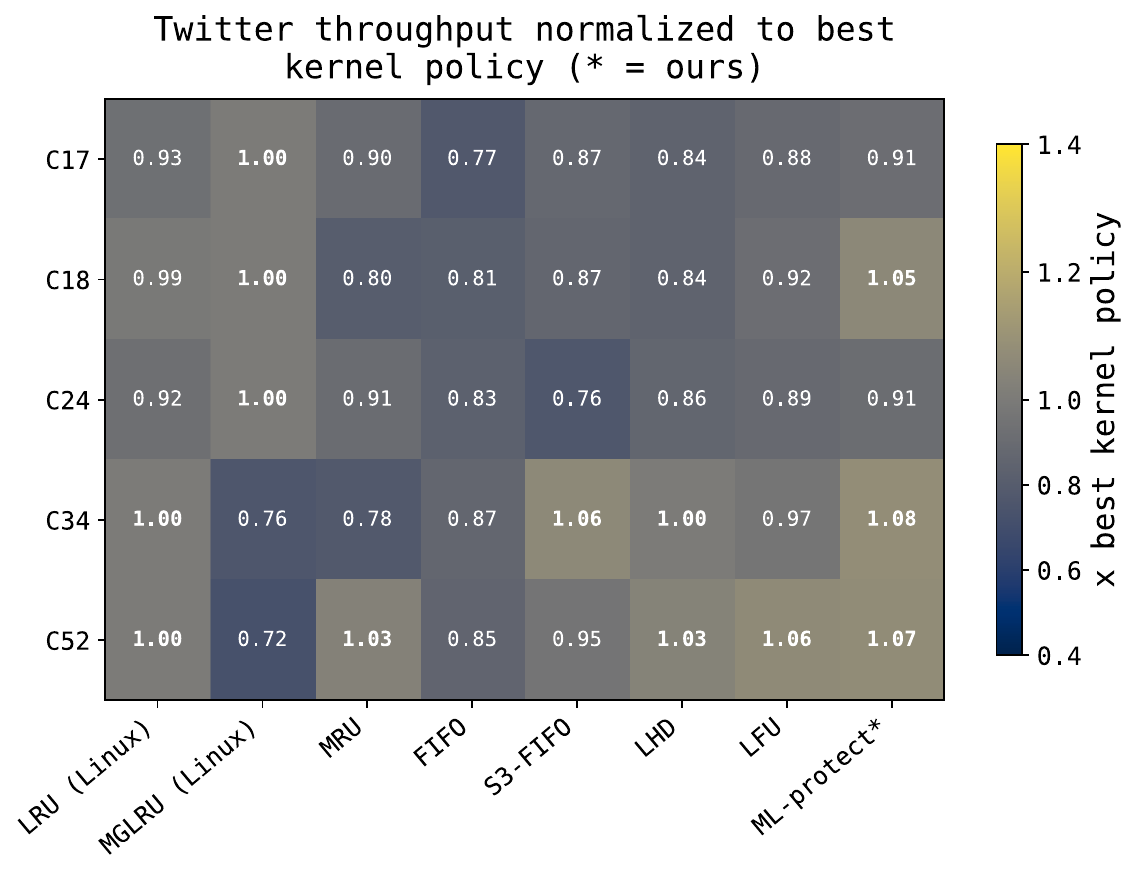}
\Description{Heatmap of Twitter cluster throughput per policy normalized to the best kernel policy.}
\caption{Twitter cluster throughput normalized to the best kernel policy
($\times$\,best kernel; ${}^{*}=$\,ours).}
\label{fig:twitter}
\end{figure}

Figures~\ref{fig:twitter-tput} and~\ref{fig:twitter} show that the model can
detect a reuse signal even in real-world workloads, tying with or beating each of
the best \code{cache\_ext} heuristics. MGLRU beats \code{ml\_protect} on Clusters 17 and 24, while
\code{ml\_protect} wins on Clusters 18, 34, and 52.
It is worth noting that in these high-pressure traces, minute differences in
eviction overhead affect throughput, which likely explains why
\code{cache\_ext}- and eBPF-implemented policies, including \code{ml\_protect}, do
not perform as well on average as the kernel built-ins.

\section{Reflections and Conclusion}
LearnedCache demonstrates that simple neural prediction algorithms can be
implemented in settings as resource-constrained as the Linux page cache using
eBPF and existing frameworks such as \code{cache\_ext}. We conclude that
perceptron-based policies are able to generalize relatively well to reproducible workloads,
at the cost of decision overhead during model inference. These findings open new
questions for future work, the most important being:
\begin{itemize}[leftmargin=*]
\item Can an ML-based policy be trained on a specific section of a real-world
workload and still perform well throughout the duration of a continuous
workload, either through reinforcement learning or further model improvements?
\item Is it possible to implement performant ML-based policies for other kernel
components using eBPF, such as the task scheduler with \code{sched\_ext}~\cite{corbet_schedext}?
\end{itemize}

We believe that LearnedCache provides a novel solution for addressing the
bottleneck caused by heuristic policies, and takes a step toward more dynamic
ML-based components in kernel space, potentially applicable to a wide range of
devices worldwide.

\bibliographystyle{ACM-Reference-Format}
\bibliography{references}

@article{stonebraker1981,
  author    = {Stonebraker, Michael},
  title     = {Operating System Support for Database Management},
  journal   = {Communications of the ACM},
  volume    = {24},
  number    = {7},
  pages     = {412--418},
  year      = {1981},
  publisher = {Association for Computing Machinery},
  doi       = {10.1145/358699.358703}
}

@misc{corbet2021mglru,
  author       = {Corbet, Jonathan},
  title        = {The Multi-Generational {LRU}},
  howpublished = {LWN.net},
  year         = {2021},
  note         = {\url{https://lwn.net/Articles/851184/}}
}

@inproceedings{cache_ext,
  author    = {Zussman, Tal and Zarkadas, Ioannis and Carin, Jeremy and Cheng, Andrew and Franke, Hubertus and Pfefferle, Jonas and Cidon, Asaf},
  title     = {cache\_ext: Customizing the Page Cache with {eBPF}},
  booktitle = {Proceedings of the ACM SIGOPS 31st Symposium on Operating Systems Principles (SOSP '25)},
  pages     = {462--478},
  year      = {2025},
  publisher = {Association for Computing Machinery},
  doi       = {10.1145/3731569.3764820}
}

@article{ml_clock,
  author    = {Cho, Minju and Kang, Dong Hyun},
  title     = {{ML-CLOCK}: Efficient Page Cache Algorithm Based on Perceptron-Based Neural Network},
  journal   = {Electronics},
  volume    = {10},
  number    = {20},
  pages     = {2503},
  year      = {2021},
  publisher = {MDPI},
  doi       = {10.3390/electronics10202503}
}

@misc{yang2023_learned_cache,
  author       = {Yang, Dongsheng and Berger, Daniel S. and Li, Kai and Lloyd, Wyatt},
  title        = {A Learned Cache Eviction Framework with Minimal Overhead},
  year         = {2023},
  eprint       = {2301.11886},
  archivePrefix = {arXiv},
  primaryClass = {cs.OS},
  note         = {\url{https://arxiv.org/abs/2301.11886}}
}

@inproceedings{song2023_halp,
  author    = {Song, Zhenyu and Chen, Kevin and Sarda, Nikhil and Alt{\i}nb{\"u}ken, Deniz and Brevdo, Eugene and Coleman, Jimmy and Ju, Xiao and Jurczyk, Pawel and Schooler, Richard and Gummadi, Ramki},
  title     = {{HALP}: Heuristic Aided Learned Preference Eviction Policy for {YouTube} Content Delivery Network},
  booktitle = {20th USENIX Symposium on Networked Systems Design and Implementation (NSDI '23)},
  year      = {2023},
  publisher = {USENIX Association}
}

@inproceedings{beckmann2018lhd,
  author    = {Beckmann, Nathan and Chen, Haoxian and Cidon, Asaf},
  title     = {{LHD}: Improving Cache Hit Rate by Maximizing Hit Density},
  booktitle = {15th USENIX Symposium on Networked Systems Design and Implementation (NSDI '18)},
  pages     = {389--403},
  year      = {2018},
  publisher = {USENIX Association}
}

@misc{gbadamosi2024ebpfruntimelinuxkernel,
  author       = {Gbadamosi, Bolaji and others},
  title        = {The {eBPF} Runtime in the Linux Kernel},
  year         = {2024},
  eprint       = {2410.00026},
  archivePrefix = {arXiv},
  primaryClass = {cs.OS},
  note         = {\url{https://arxiv.org/abs/2410.00026}}
}

@book{silberschatz2018operating,
  author    = {Silberschatz, Abraham and Galvin, Peter B. and Gagne, Greg},
  title     = {Operating System Concepts},
  edition   = {10th},
  year      = {2018},
  publisher = {Wiley}
}

@misc{corbet2021folios,
  author       = {Corbet, Jonathan},
  title        = {Clarifying Memory Management with Page Folios},
  howpublished = {LWN.net},
  year         = {2021},
  note         = {\url{https://lwn.net/Articles/849538/}}
}

@misc{linux_folio_mark_accessed,
  author       = {{The Linux Kernel Developers}},
  title        = {\texttt{folio\_mark\_accessed}},
  howpublished = {Linux kernel source, \texttt{mm/swap.c}},
  year         = {2024},
  note         = {\url{https://elixir.bootlin.com/linux/latest/source/mm/swap.c}}
}

@misc{ebpf_io,
  author       = {{The eBPF Authors}},
  title        = {{eBPF} Documentation},
  howpublished = {\url{https://ebpf.io/what-is-ebpf/}},
  year         = {2024}
}

@misc{leveldb,
  author       = {{Google}},
  title        = {{LevelDB}},
  howpublished = {\url{https://github.com/google/leveldb}}
}

@inproceedings{yang2020twitter,
  author    = {Yang, Juncheng and Yue, Yao and Rashmi, K. V.},
  title     = {A Large Scale Analysis of Hundreds of In-Memory Cache Clusters at {Twitter}},
  booktitle = {14th USENIX Symposium on Operating Systems Design and Implementation (OSDI 20)},
  pages     = {191--208},
  year      = {2020},
  publisher = {USENIX Association}
}

@inproceedings{yang2023s3fifo,
  author    = {Yang, Juncheng and Zhang, Yazhuo and Qiu, Ziyue and Yue, Yao and Vinayak, Rashmi},
  title     = {{FIFO} Queues Are All You Need for Cache Eviction},
  booktitle = {Proceedings of the 29th Symposium on Operating Systems Principles (SOSP '23)},
  pages     = {130--149},
  year      = {2023},
  publisher = {Association for Computing Machinery}
}

@misc{corbet_schedext,
  author       = {Corbet, Jonathan},
  title        = {The Extensible Scheduler Class},
  howpublished = {LWN.net},
  note         = {\url{https://lwn.net/Articles/922405/}}
}

@misc{cgroupv2,
  author       = {Heo, Tejun},
  title        = {Control Group v2},
  howpublished = {Linux kernel documentation},
  note         = {\url{https://docs.kernel.org/admin-guide/cgroup-v2.html}}
}

@misc{linux_filemap,
  author       = {{The Linux Kernel Developers}},
  title        = {\texttt{\_\_filemap\_add\_folio} / \texttt{\_\_filemap\_remove\_folio}},
  howpublished = {Linux kernel source, \texttt{mm/filemap.c}},
  year         = {2024},
  note         = {\url{https://elixir.bootlin.com/linux/latest/source/mm/filemap.c}}
}

@article{oneil1996lsm,
  author    = {O'Neil, Patrick and Cheng, Edward and Gawlick, Dieter and O'Neil, Elizabeth},
  title     = {The Log-Structured Merge-Tree ({LSM}-Tree)},
  journal   = {Acta Informatica},
  volume    = {33},
  number    = {4},
  pages     = {351--385},
  year      = {1996},
  publisher = {Springer},
  doi       = {10.1007/s002360050048}
}

@article{wisckey,
  author    = {Lu, Lanyue and Pillai, Thanumalayan Sankaranarayana and Gopalakrishnan, Hariharan and Arpaci-Dusseau, Andrea C. and Arpaci-Dusseau, Remzi H.},
  title     = {{WiscKey}: Separating Keys from Values in {SSD}-Conscious Storage},
  journal   = {ACM Transactions on Storage},
  volume    = {13},
  number    = {1},
  year      = {2017},
  publisher = {Association for Computing Machinery}
}

\end{document}